# Excitation of fundamental shear horizontal wave by using face-shear ($d_{36}$) piezoelectric ceramics

Hongchen Miao[1,2], Shuxiang Dong[3], Faxin Li[1,2,a]

[1] LTCS and Department of Mechanics and Engineering Science, College of Engineering, Peking University, Beijing, 100871, China

[2] Center for Applied Physics and Technology, Peking University, Beijing, 100871, China

[3] Department of Material Science and Engineering, College of Engineering, Peking University, Beijing, 100871, China

**Abstract**

The fundamental shear horizontal (SH0) wave in plate-like structures is extremely useful for non-destructive testing (NDT) and structural health monitoring (SHM) as it is non-dispersive. However, currently the SH0 wave is usually excited by electromagnetic acoustic transducers (EMAT) whose energy conversion efficiency is fairly low. The face-shear ($d_{36}$) mode piezoelectrics is more promising for SH0 wave excitation but this mode cannot appear in conventional piezoelectric ceramics. Recently, by modifying the symmetry of poled $PbZr_{1-x}Ti_xO_3$ (PZT) ceramics via ferroelastic domain engineering, we realized the face-shear $d_{36}$ mode in both soft and hard PZT ceramics. In this work, we further improved the face-shear properties of PZT-4 and PZT-5H ceramics via lateral compression under elevated temperature. It was found that when bonded on a 1 mm-thick aluminum plate, the $d_{36}$ type PZT-4 exhibited better face-shear performance than PZT-5H. We then successfully excite SH0 wave in the aluminum plate using a face-shear PZT-4 square patch and receive the wave using a face-shear PMN-PT patch. The frequency response and directionality of the excited SH0 wave were also investigated. The SH0 wave can be dominate over the Lamb waves (S0 and A0 waves) from 160 kHz to 280 kHz. The wave amplitude reaches its maxima along the two main directions (0° and 90°). The amplitude can keep over 80% of the maxima when the deviate angle is less than 30°, while it vanishes quickly at the 45 ° direction. The excited SH0 wave using piezoelectric ceramics could be very promising in the fields of NDT and SHM.

---

[a] Author to whom all correspondence should be addressed, Email: lifaxin@pku.edu.cn

**1. Introduction**

In the past decades, the guided wave method has become more and more important in the field of non-destructive testing (NDT) and structural health monitoring (SHM) as this type wave is less dissipative and suitable for long-distance inspection.[1] In general, there are two different families of guided waves that can exist in a plate-like structure: Lamb waves and shear horizontal (SH) waves. Compared with the dispersive Lamb waves, the fundamental shear horizontal (SH0) wave is of great practical importance due to its unique features.[2,3] Firstly, SH0 wave is non-dispersive, namely, its phase and group velocities are not frequency-dependent, which can simplify the interpretation of signals. Secondly, SH0 wave is less affected by the presence of surrounding media, since there is no out-of-plane particle displacement in this wave mode. Furthermore, SH0 wave will not convert to Lamb waves at defects or boundaries, reducing the complexity of the received signals.

In spite of above-mentioned attractive features of the SH0 wave, typically it is not straightforward to excite this wave. In the past decades, several methods have been proposed to excite the SH0 waves among which the electromagnetic acoustic transducer (EMAT) method is the well known solution. There are two kinds of EMATs used for SH wave excitation: periodic permanent magnet (PPM) EMATs and magnetostrictive EMATs.[2] The PPM EMATs are based on the Lorentz force,[4] while the magnetostrictive EMATs are based on magnetostrictive effects.[5] However, both EMATs can only be used for conductive metallic structures. Furthermore, the EMATs are non-contact transducers and their energy conversion efficiency is fairly low compared with the contact piezoelectric counterpart, leading to a lower signal-to-noise ratio (SNR). Therefore, the EMATs are usually used to detect structures with coatings or under high temperature in which the piezoelectric transducers are not applicable. In addition, the EMATs require high power excitation, which is not suitable for SHM applications.

Piezoelectric transducers have been widely used in both NDT and SHM fields, while it is not easy to generate SH0 wave by using conventional piezoelectric transducers.[6] Kamal et al. used a thickness-shear ($d_{15}$) type piezoelectric patch to generate SH0 wave perpendicular to the poling direction.[7] However, strong lamb waves were also excited simultaneously along the poling direction. The amplitude of Lamb waves can be reduced by optimizing the geometry of thickness-shear type piezoelectric patch.[8] Unfortunately, the first resonance frequency of the $d_{15}$ piezoelectric patches is very high (about 1 MHz), which approaches or even exceeds the cut off frequency of the SH1 wave in a 2 mm-thick aluminum plate. Therefore, the amplitude of the excited SH0 waves is usually very small (typically of only several millivolts),[7] since the deformation of the $d_{15}$ piezoelectric patches cannot be amplified by resonance. In 2005, a new face-shear ($d_{36}$) mode was realized in [011]-poled rhombohedral (1-x)[Pb(Mg$_{1/3}$Nb$_{2/3}$)O$_3$]-x[PbTiO$_3$] (PMN-PT) crystals with $Zxt \pm 45°$ cut direction,[9] but this mode cannot appear in conventional piezoelectric ceramics because of its transversally isotropic symmetry. Recently, Zhou et al. successfully generated and received the SH0 waves in an aluminum plate using the $d_{36}$ type PMN-PT patches.[10,11] The face-shear ($d_{36}$) mode is superior to the thickness-shear ($d_{15}$) mode, since its working voltage is along the poling direction. However, such face-shear PMN-PT piezoelectric crystals cannot be widely used for NDT or SHM because of its lower Curie temperature, less stable domains and high cost. To solve these challenges, we recently realized the face-shear ($d_{36}$) mode in both hard and soft PbZr$_{1-x}$Ti$_x$O$_3$ (PZT) ceramics via ferroelastic domain engineering.[12,13]

In this work, we furtherly improved the face-shear properties of PZT ceramics including PZT-4 and PZT-5H via compression induced ferroelastic domain switching at elevated temperature. The piezoelectric coefficient $d_{36}$ and electromechanical coupling factor $k_{36}$ were enhanced significantly, which is important for generating SH0 waves. Then the face-shear PZT patches were bonded on a 1mm-thick aluminum plate to excite SH0 waves and the face-shear PMN-PT patches were used to receive the waves. Results show that the SH0 wave was successfully excited with high signal-to-noise ratio (SNR). The frequency responses and directionality of the excited SH0 wave were also studied.

## 2. Experimental methods

### 2.1 Fabrication of $d_{36}$ type PZT ceramics

Conventional PZT-4 and PZT-5H piezoelectric ceramics, which are provided by the Institute of Acoustics, Chinese Academy of Sciences, are used here to fabricate the $d_{36}$ type ceramics. The material parameters have been provided in our previous work[13] and will not be listed here. These PZT ceramics were firstly cut into cube-shaped samples ($9\,\text{mm} \times 9\,\text{mm} \times 9\,\text{mm}$ for PZT-4, $8\,\text{mm} \times 8\,\text{mm} \times 8\,\text{mm}$ for PZT-5H) for lateral compression. The temperature-controlled compression testing setup is illustrated in Fig. 1 (a). The cube-shaped PZT sample is immersed in an oil tank filled with silicon oil whose boiling point is about 200 ℃. A temperature control system consisting of a metal heater and a thermocouple sensor was used to control the temperature of the silicon oil with the resolution of 0.2 ℃. Lateral compressive stress ($T_2$) perpendicular to the polar axis was applied to the samples by using a material testing machine (WDW-100, Changchun material testing machine Ltd, China). In order to avoid any possible bias compression, a loading head with a spherical hinge was used. Two alumina blocks with dimensions of $25 \times 25 \times 10\,\text{mm}^3$ were employed to insulate the PZT specimen from the loading head, as shown in Fig. 1(a).

The lateral compression testing was conducted as follow. Firstly, a small preload of ~1MPa was applied to the specimen to maintain the contact. The specimen was then heated at the rate of about 3 ℃/min to the target temperature (25 ℃ and 80 ℃ for PZT-5H, 25 ℃ and 110 ℃ for PZT-4). It should be noted that the target temperature should be considerably lower than the Curie temperature to avoid thermal depolarization. Then the target temperature was holding and the compressive stress was gradually applied to the specimen until the maximum stress was reached ($180\,\text{MPa}$ for PZT-5H and $300\,\text{MPa}$ for PZT-4).The maximum stress was kept for two hours to make the switched domains more stable. Later, the PZT specimen was gradually cooled to room temperature and the maximum compressive stress was still held during cooling. Then the compressive stress was removed gradually with the unloading rate of about $0.5 \text{ MPa/s}$. After compression, the pseudocrystal symmetry of the poled PZT samples was changed from transversally isotropic to be orthogonal, resulting in that its $d_{31}$ is larger than $d_{32}$. Finally, the compressed samples were cut along the $Zxt \pm 45°$ direction, as introduced in our recent work.[12] Fig. 1(b) shows the photo of a $Zxt \pm 45°$ cut PZT sample. The theoretical $d'_{36}$ in the cut sample can be obtained by[12]

$$d'_{36} = \pm(d_{32} - d_{31}) \qquad (1).$$

Thereafter, the cut samples were sliced into thin square patches ($6.3 \text{ mm} \times 6.3 \text{ mm} \times 1 \text{ mm}$ for PZT-4, $5.6 \text{ mm} \times 5.6 \text{ mm} \times 1 \text{ mm}$ for PZT-5H) for impedance measurement and SH0 wave excitation. The impedance spectra of the PZT patches are measured by an impedance analyzer (HP4294A, Agilent Technologies). The detailed method of measuring the face-shear piezoelectric properties of these PZT ceramics was the same as that introduced in our recent work.[13]

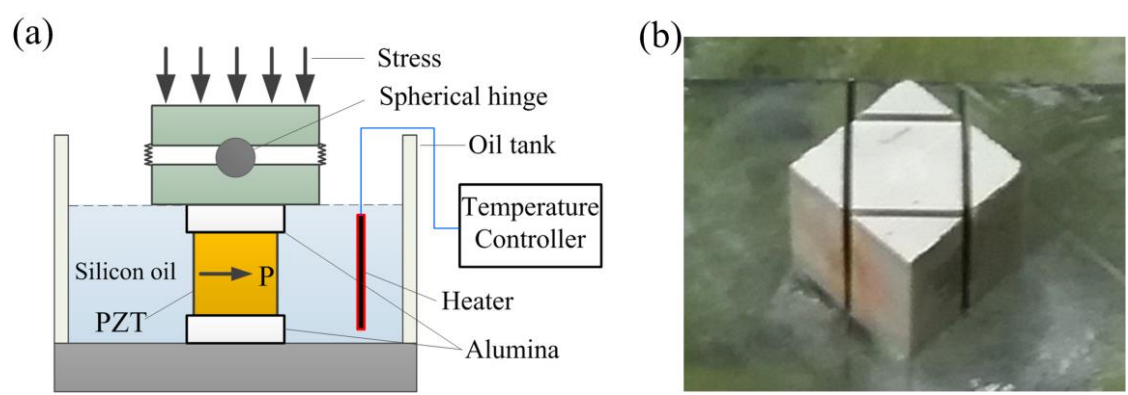

Fig 1. (a) Temperature controlled compression setup for fabricating the face-shear ($d_{36}$) type PZT ceramics, (b) the photo of a $Zxt \pm 45°$ cut PZT sample.

## 2.2 Excitation of SH0 waves using $d_{36}$ type PZT patches

For the SH0 wave excitation, here an aluminum plate with the dimensions of $1000 \text{ mm} \times 1000 \text{ mm} \times 1 \text{ mm}$ was employed. It is well known that hard PZT is more suitable for actuator because of its high mechanical quality factor, high resistance to depolarization and low dielectric losses. To explore which type of face-shear PZT ceramic (PZT-4 and PZT-5H) is more suitable for generating SH0 waves, different PZT patches ($6.3 \text{ mm} \times 6.3 \text{ mm} \times 1 \text{ mm}$ for PZT-4, $5.6 \text{ mm} \times 5.6 \text{ mm} \times 1 \text{ mm}$ for PZT-5H) were firstly bonded on the surface of an aluminum plate with a conductive glue and the impedance spectra were measured to check their face-shear performance. After that, the PZT patch with better face-shear performance was selected as the SH0 wave actuator. As introduced in our previous study,[12] the extensional mode ($d_{31}$) always co-existed with the face-fear mode ($d_{36}$) in face-shear type PZT ceramics and the $d_{31}$ mode is dominant at most frequencies. The pure $d_{36}$ mode can only be obtained at its resonance frequency and the bandwidth of the $d_{36}$ resonance is very narrow. Thus, the $d_{36}$ type PZT ceramics is not a good candidate sensor for SH wave reception. Here we employed the $d_{36}$ type PMN-PT (0.72[Pb(Mg$_{1/3}$Nb$_{2/3}$)O$_3$]-0.28[PbTiO$_3$]) crystal as sensors to receive SH0 waves, since its piezoelectric coefficient $d_{36}$ (~1600 pC/N) is much larger than $d_{31}$ (~-360 pC/N).[14] In practical applications, the SH0 wave can also be detected by other transducers such as the fibre Bragg grating sensors.[15] To investigate the directionality of the generated SH0 waves, four PMN-PT sensors with dimension of $5 \text{ mm} \times 5 \text{ mm} \times 1 \text{ mm}$ were arranged around the PZT actuator along the 0°, 15°, 30° and 45° direction respectively. The layout and location of the actuator and sensors are shown in Fig. 2(a). The PZT actuator was driven by a five-cycle Hanning window-modulated sinusoid toneburst signal provided by a function generator (3320A, Agilent, USA). The amplitude of the drive signal is amplified by a power amplifier (KH7602M) to 40 V. An Agilent DSO-X 3024A oscilloscope was used to collect the wave signals received by the PMN-PT sensors. The SH0, A0 and S0 wave modes are identified based on the their different group velocities. The group velocity dispersion curves of S0 and A0 waves can be easily calculated by using the software developed by Professor Giurgiutiu's group in University of South Carolina, US (http://www.me.sc.edu/research/lamss/html/software.html.). As for the SH0 wave, its group velocity is equal to the bulk shear wave speed. The calculated group velocity dispersion curves of these three wave modes in the 1 mm-thick aluminum plate are plotted in Fig. 2(b).

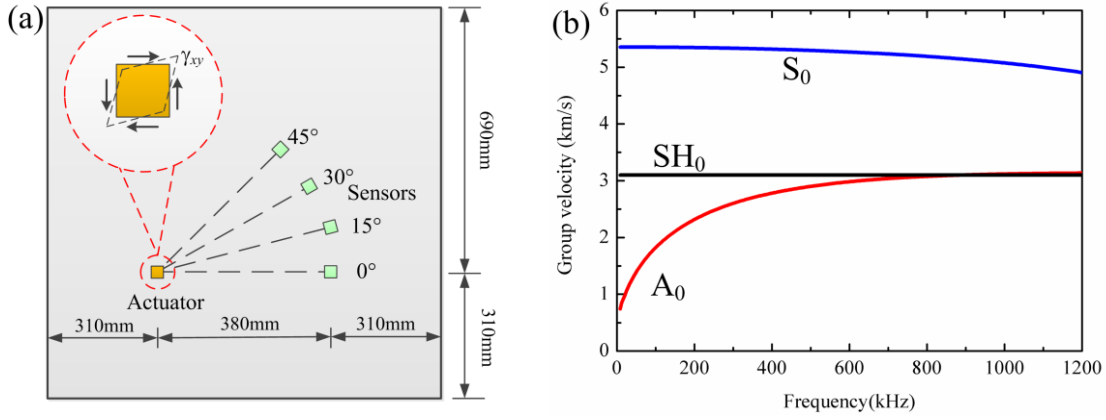

Fig. 2 (a) Schematic of the layout of piezoelectric actuator and sensors for guided wave generation and detection in a 1 mm-thick aluminum plate; (b) group velocity curves of the Lamb wave and SH0 waves in the plate.

## 3. Results and Discussions

### 3.1 Face-shear performances of the $d_{36}$ type PZT patches

Fig. 3 shows the measured impedance spectra of the face-shear ($d_{36}$) mode in the PZT-5H and PZT-4 patches. It can be seen that the face-shear resonance and anti-resonance peaks of both PZT-5H and PZT-4 patches fabricated under elevated temperature are more pronounced than that of those fabricated at room temperature (25 ℃). When the testing frequency varies from resonance to anti-resonance, the impedances change in the $d_{36}$ type PZT-5H ceramics fabricated at 25 ℃ is 219 Ohm, while this value increases to 1055 Ohm for that fabricated at 80 ℃. Similar phenomenon was also observed in the $d_{36}$ type PZT-4 ceramics. On the other hand, it can be seen from Fig. 3 that the $d_{36}$ mode resonance frequencies of both PZT-5H and PZT-4 patches fabricated under elevated temperature are higher than that of those fabricated at room temperature. This tendency is more significant in the $d_{36}$ type PZT-4 ceramics, as shown in Fig. 3(b). The measured face-shear properties of the $d_{36}$ type PZT samples fabricated under different temperatures were listed in Table Ⅰ. The elevated-fabrication-temperature induced changes in frequency constants $N_{36}$ can be attributed to the decrease of the elastic constant $s_{66}^{E}$ of PZT ceramics. The domain configurations in PZT samples fabricated under elevated temperature are more close to the two-dimensional domain states, since more ferroelastic domains can be switched by the lateral compression under high temperature. The elastic constant $s_{66}^E$ for the two-dimensional domain states is usually smaller than that for three-dimensional domain states. Similar phenomenon has also been reported in single crystals.[16]

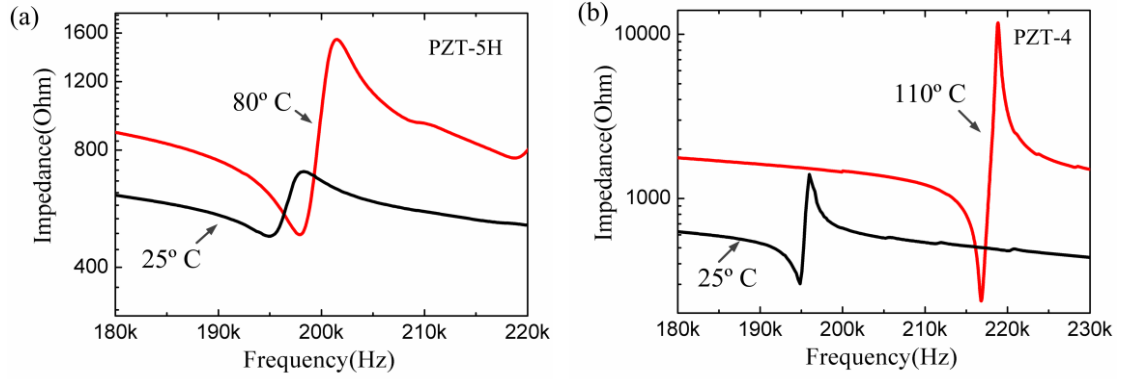

Fig. 3 Impedance spectra of the face-shear ($d_{36}$) mode in the (a) PZT-5H patch ($5.6 \text{ mm} \times 5.6 \text{ mm} \times 1 \text{ mm}$) and (b) PZT-4 patch ($6.3 \text{ mm} \times 6.3 \text{ mm} \times 1 \text{ mm}$) fabricated under different temperature conditions.

From the measured material parameters listed in Table I, it can be seen that the face-shear performances of PZT-4 ceramics fabricated at 110 °C were significantly improved when compared with that fabricated at room temperature. The piezoelectric coefficient $d_{36}$ of PZT-4 ceramics was improved from 93 pC/N to 108 pC/N and the electromechanical coupling factor $k_{36}$ increased from 0.14 to 0.19. Note that these two parameters are extremely important for actuator applications. The improvement of face-shear properties of PZT-4 ceramics was attributed to that more ferroelastic domains switched under compression at high temperature, as our previous investigations show that at room temperature ferroelastic domain switching in PZT-4 ceramics cannot saturate even under the lateral compression of 300 MPa.[13] In comparison, little change were observed in the $d_{36}$ and $k_{36}$ of PZT-5H ceramics when fabricated at elevated temperature, which may imply that even at room temperature ferroelastic domain switching in PZT-5H is nearly saturated under the compressive stress of 180 MPa.[13]

On the other hand, the mechanical quality factor $Q_{36}$ and the elastic constant $s_{66}^E$ are also very important for face-shear actuators. It can be seen from Table I that when fabricated at elevated temperature, the PZT-4 ceramics show little change in $Q_{36}$ but the material become stiffer than that fabricated at room temperature. In comparison, both parameters of PZT-5H are not sensitive to the fabrication temperature. Note that although the face-shear PZT-4 ceramics always has a smaller $d_{36}$ and $k_{36}$ than PZT-5H, it is always stiffer and has a larger $Q_{36}$. Thus, it cannot be determined from Table I whether the face-shear PZT-4 or PZT-5H is more suitable for actuator applications.

TABLE Ⅰ. The measured face-shear properties of the $d_{36}$ type PZT samples fabricated under different temperatures.

|  | Fabrication temperature | $d_{36}$ (pC/N) | $k_{36}$ | $Q_{36}$ | $N_{36}$ (Hzm) | $s_{66}^E$ (pm$^2$/N) |
|---|---|---|---|---|---|---|
| PZT-4 | 25 ℃ | 93 | 0.14 | 138 | 1227 | 34.70 |
|  | 110 ℃ | 108 | 0.19 | 136 | 1366 | 30.55 |
| PZT-5H | 25 ℃ | 273 | 0.23 | 64 | 1092 | 43.94 |
|  | 80 ℃ | 274 | 0.24 | 60 | 1108 | 42.64 |

To further check the face-shear performance of the $d_{36}$ type PZT-4 and PZT-5H ceramics fabricated under elevated temperature, we bonded both square patches on a 1 mm-thick aluminum plate and measured their impedance spectra, as shown in Fig. 4, where the impedance spectra (Fig. 4(a) and Fig. 4(b)) of both free patches were also plotted for comparison. It can be seen from Fig. 4(c) and Fig. 4(d) that when the $d_{36}$ type PZT patches were mounted on an aluminum plate, the resonance modes changed significantly. The $d_{36}$ resonance peak can still be observed in the bonded PZT-4 patch at about 255 kHz in Fig. 4(c), but the peak amplitude is greatly reduced compared to the case of free resonance mode in Fig. 4(a). In comparison, the $d_{36}$ resonance peak of the PZT-5H patch vanishes when it is bonded on the plate. This may imply that the constraint by the aluminum plate is too strong for the PZT-5H patch thus the face-shear mode is suppressed. Therefore, we think the $d_{36}$ type PZT-4 patch is more suitable for generating SH0 waves in thin plates.

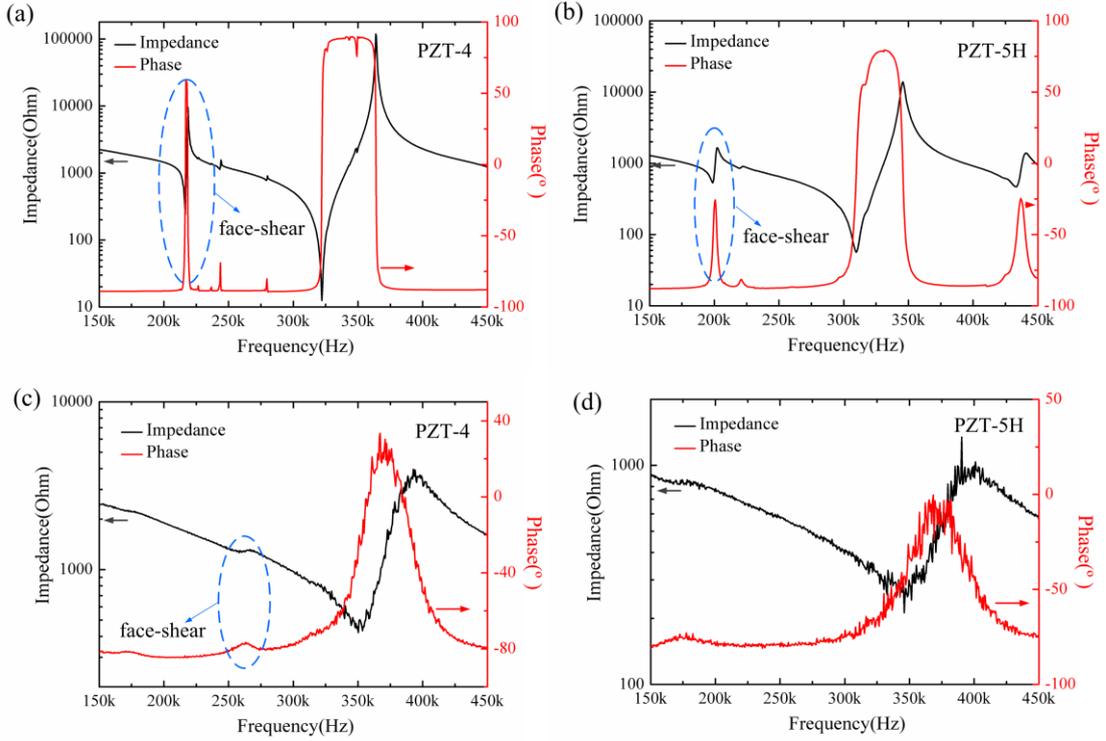

Fig. 4 The impedance spectra of the free $d_{36}$ type PZT square patches ((a) and (b)) and the $d_{36}$ type PZT patches bonded on a 1 mm-thick aluminum plate ((c) and (d)).

**3.2 Properties of SH0 waves excited by a $d_{36}$ type PZT-4 patch**

a) Identification of the excited SH0 wave

Here a face-shear type PZT-4 patch with the dimension of $6.3\ \text{mm} \times 6.3\ \text{mm} \times 1\ \text{mm}$ was used to excite the SH0 waves. The central frequency of the excitation signal was firstly set at 255 kHz, which is near the face-shear resonance frequency of the bonded PZT-4 patch as shown in Fig. 4(c). The wave signals received by the PMN-PT sensor at 0° were shown in Fig. 5(a). It can be seen that SH0 wave with high SNR (signal-to-noise ratio) was excited successfully by using the face-shear type PZT-4 patch. Fig. 5(b) shows the wave signals after continuous wavelet transform (CWT) from which it can be seen that the time interval between the peak of the driving signal and that of the received signal is 125.48 μs. Bearing in mind the distance between actuator and sensor is 380 mm, the group velocity is then calculated to be $3028 \text{ m/s}$, which is in good agreement with the theoretical group velocity of SH0 wave ($3099 \text{ m/s}$) in this aluminum plate. It should be noted that, although the $d_{36}$ type PZT-4 patch was excited near the resonance frequency of the face-shear mode, the S0 and A0 wave modes were also excited simultaneously as shown in Fig. 5(a), which is attributed to the co-existed $d_{31}$ mode in the $d_{36}$ type PZT-4 patches. Such phenomenon was also observed in the face-shear type PMN-PT single crystals, using which pure SH0 wave is still difficult to be excited.[11] Fortunately, the amplitude of the excited SH0 wave is obviously larger than that of S0 and A0 waves, as shown in Fig. 5. Moreover, the S0 and A0 waves will become attenuated in amplitude during propagation, since both waves are dispersive.

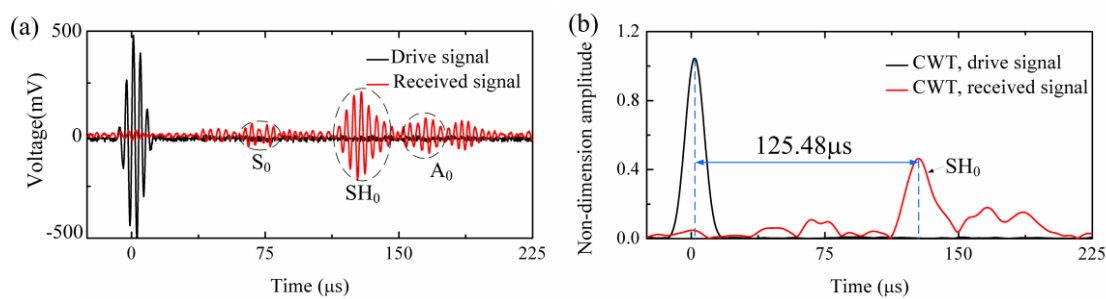

Fig. 5 (a) wave signals excited by the face-shear PZT-4 patch at 255 kHz and received by the PMN-PT sensor placed at 0° direction; (b) wave signals analyzed by using continuous wavelet transform (CWT).

b) Frequency responses of the excited SH0 wave

Fig. 6 shows the wave signals excited at different central frequencies and received by the PMN-PT sensor placed at 0 ° direction. It can be seen that the tuning frequency characteristics are different for SH0, S0 and A0 wave modes. When the excited central frequency varies from 100 kHz to 300 kHz, the SH0 wave mode is dominant from about 160 kHz to 280 kHz. The amplitude of the SH0 wave reaches its maxima around 220 kHz, then decreases monotonically. For the S0 wave mode, almost no S0 wave was detected below 190 kHz, then it becomes more and more stronger with the increasing driving frequency and becomes dominant around 300 kHz. As to the A0 wave mode, it reaches the first amplitude peak at about 220 kHz, then deceases to the minimum amplitude at about 280 kHz and then increases again. The tuning frequency characteristics of these guided wave modes were attributed to that the deformation of the $d_{36}$ type PZT-4 patch is strongly frequency dependent. Note that the extensional $d_{31}$ mode always co-existed with the $d_{36}$ mode in $d_{36}$ type PZT ceramics. Thus the field induced deformation of $d_{36}$ type PZT-4 patch is the superposition of an extensional deformation and a face-shear deformation. When the excited frequency is near the resonance frequency of the $d_{36}$ mode, the SH0 wave will be dominate. Similarly, when the excited frequency is near the resonance frequency of the $d_{31}$ mode, the S0 and A0 waves will become dominate. On the other hand, the amplitude of the excited waves will reach its maxima when the length of the PZT patch equals to odd multiple of the half wavelength. Also its amplitude reaches the minima when the length of the PZT patch is a multiple of the wavelength. Since the wavespeed and wavelength of S0, A0 and SH0 waves are different, such matching between the PZT patch length and wavelength will occur at different frequencies for different wave modes.[17] By careful examination, it can also be seen from Fig. 6 that both S0 and A0 waves are dispersive as their waveform and group velocity are strongly influenced by the excitation frequency. Also as expected, the group velocity of the excited SH0 wave is frequency-independent, as shown clearly in Fig. 6.

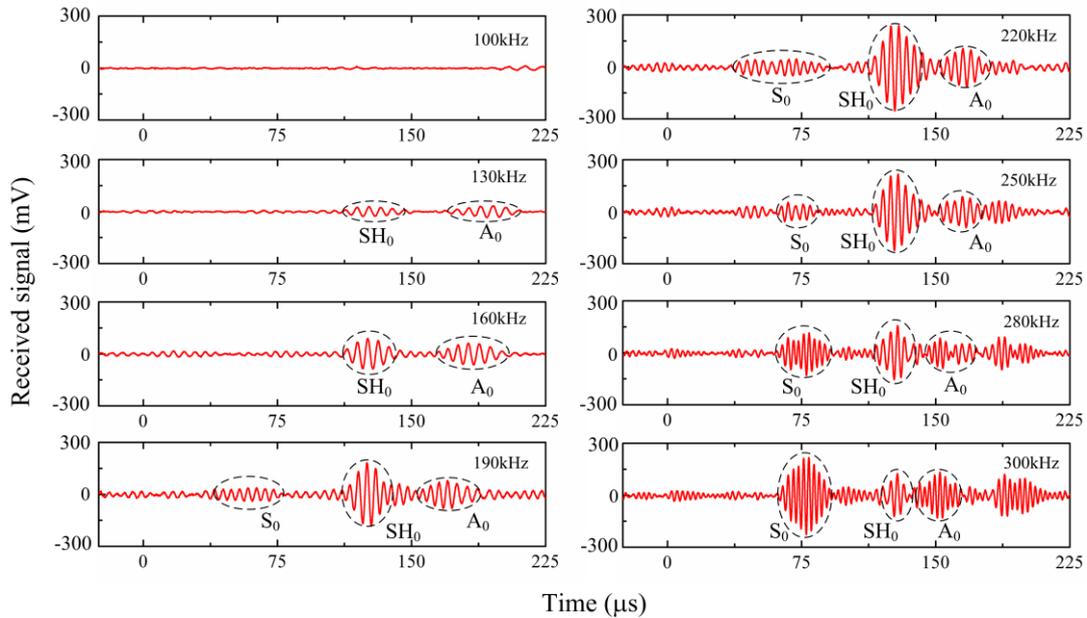

Fig. 6 Wave signals excited by the face-shear PZT-4 patch at different central frequencies and received by the PMN-PT sensor placed at 0 ° direction.

c) Directionality of the excited SH0 wave

Fig. 7(a) shows the amplitudes of the SH0 waves excited at 160 kHz and received along different propagation directions. The reason that the central frequency was fixed at 160 kHz is that the S0 wave mode disappeared at this frequency. It can be seen that the amplitude of the excited SH0 wave reaches its maximum value when detected along the 0° direction. The amplitude of the SH0 wave decreases to about 90% of its maxima at 15° direction and then further decreases to about 80% of its maxima at 30° direction. At 45° direction, almost no SH0 wave was detected. Note that the similar phenomenon was also observed in the face-shear type PMN-PT crystals.[11] Such phenomenon is not difficult to understand. The pure face-shear deformation ($S_6$) is equivalent to the superposition of an elongation (or contraction) deformation along the 45° direction ($S_{45°}$) and a contraction (elongation) deformation along the 135° direction ($S_{135°}$), as shown in Fig. 7(b). Therefore, the alternating elongation/contraction deformation along 45° direction will generate lamb waves. The excited SH0 wave is symmetric along the 45° direction, i.e., the amplitude will increase when the deviate angle is above 45° and reaches its maxima again at 90°.

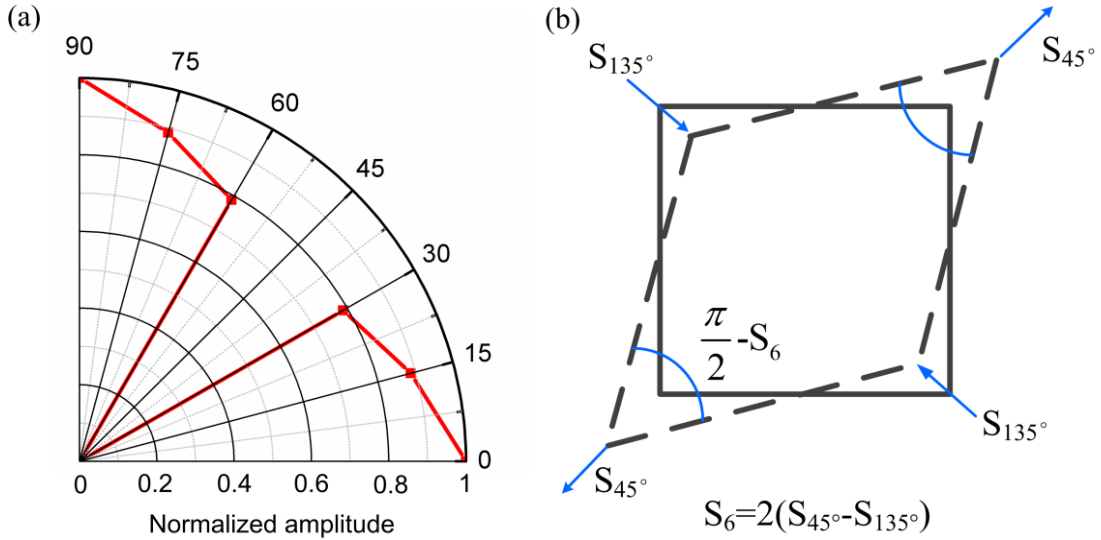

Fig. 7 (a) Normalized amplitude of SH0 wave excited at 160 kHz along different propagation directions, (b) schematic of the face-shear deformation.

## 4. Conclusion

In summary, we realized and improved the face-shear ($d_{36}$) properties of PZT ceramics via stress induced ferroelastic domain switching under elevated temperature, and successfully excited the SH0 wave in an aluminum plate using the $d_{36}$ type PZT-4 patches. Over a wide frequency range (160 kHz to 280 kHz in this work), the excited SH0 wave is dominant compared with the simultaneously excited Lamb wave (S0 and A0 waves). The directionality of the excited SH0 wave was further investigated. It was found that amplitude is maximum along the two main directions (0° and 90°), and keeps about 80% of its maxima at 30° direction and quickly vanishes at 45° direction. The excited SH0 wave by using PZT patches is of great importance to the field of NDT and SHM since the SH0 wave is non-dispersive and the PZT patch is small and cost-effective. Furthermore, as the SH0 wave in a thin plate is essentially equivalent to the fundamental torsional [T (0,1)] wave in a thin pipe, the T(0,1) wave is also expected to be excited by the face-shear PZT ceramics.